\documentclass[preprint,3p,times, twocolumn]{elsarticle}

\usepackage{graphicx}
\usepackage{color}
\usepackage{textcomp}
\usepackage[T1]{fontenc}  
\usepackage[utf8]{inputenc} 

\journal{Physics Letters B}

\begin{document}

\begin{frontmatter}

\title{Quasi-elastic scattering for the nuclear ground state structure: An intriguing case of $^{30}$Si}


\author[BARC,HBNI]{Y. K. ~Gupta\corref{cor1}}
\cortext[cor1]{Corresponding author, Email:ykgupta@barc.gov.in}

\author[GANIL]{B. Maheshwari}
\author[BARC]{G. K. Prajapati}
\author[Amity,IITR]{A. K. Jain}
\author[KYOTO]{K. Hagino}
\author[BARC,HBNI]{B. N. Joshi}
\author[BARC]{A. Pal}
\author[BARC]{N. Sirswal}
\author[BARC,HBNI]{Pawan Singh}
\author[BARC]{S. Dubey}
\author[SIES]{V. V. Desai}
\author[TIFR]{V. Ranga}
\author[BARC]{V. B. Katariya}
\author[SVNIT]{D. Patel}
\author[BARC]{H. Vyas}
\author[IITR]{S. Panwar}
\author[BARC]{B. V. John}
\author[TIFR]{I. Mazumdar}
\author[BARC]{B. K. Nayak}
\author[ND]{U. Garg}

\address[BARC]{Nuclear Physics Division, Bhabha Atomic Research Centre, Mumbai - 400085, India}
\address[HBNI]{Homi Bhabha National Institute, Anushaktinagar, Mumbai 400094, India} 
\address[GANIL]{Grand Acc\'el\'erateur National d'Ions Lourds, CEA/DSM--CNRS/IN2P3, Bvd Henri Becquerel, BP 55027, F-14076 Caen, France}
\address[Amity]{Amity Institute of Nuclear Science \& Technology, Amity University UP, NOIDA}
\address[IITR]{Department of Physics, IIT Roorkee, Roorkee-247667, India}
\address[KYOTO]{Department of Physics, Kyoto University, Kyoto 606-8502, Japan}
\address[SIES]{SIES College of Arts, Science and Commerce, Mumbai-400022, India}
\address[TIFR]{Tata Institute of Fundamental Research, Mumbai 400005, India}
\address[SVNIT]{Department of Physics, Sardar Vallabhbhai National Institute of Technology, Surat -395007, India}
\address[ND]{Department of Physics and Astronomy, University of Notre Dame, Notre Dame, IN 46556, USA}

\date{\today}

\begin{abstract}
Quasi-elastic (QEL) scattering measurements have been performed using the $^{28, 30}$Si projectiles off the $^{90}$Zr target at energies around the Coulomb barrier. Coupled-channels (CC) calculations were carried out in a large parameter space of quadrupole and  hexadecapole deformations for the N=Z, $^{28}$Si and N=Z+2, $^{30}$Si nuclei. $^{28}$Si at the N=Z line is observed to be uniquely oblate shaped in its ground state. In contrast, for $^{30}$Si with just two additional neutrons- oblate, prolate, and spherical  CC descriptions are equally compatible with the measurements.  To further investigate the nuclear structure evolution with varying neutron number, shell-model calculations were performed. These calculations reveal a sudden change in the nuclear structure aspects at $^{30}$Si  in going from $^{28}$Si to $^{30}$Si. Combined reaction and structure analyses consistently indicate that $^{30}$Si does not possess a well-defined intrinsic shape, and it is a potential candidate for ``shape fluctuations" in its ground state.
\end{abstract}

\end{frontmatter}

Investigation of nuclear shapes is a subject of prime importance because of their key roles in the study of nuclear structure \cite{Taniuchi2025}, nucleo-synthesis \cite{SchatzPR1998}, nuclear fission \cite{AndreyevReport2018, Ibrahim24}, and neutrino less double beta decay \cite{NLDBD2017}. Determination of the exact shapes of atomic nuclei within the discrete quantum states is inherently challenging. Along with low energy spectroscopic \cite{Cline_CE_1998, Yang_LASER_2023} and scattering techniques \cite{Leo_neutron_28Si,Fabrici1980, Cole1966, KokameAlpha, Siemaszko_JPG1994}, nuclear collisions at ultra-relativistic speeds are also being employed \cite{238UBeta4PRL, nature_shape_2024, RPP_RHIC_shape_2025, Hagino_RHIC_2025}. The nuclear shapes are primarily governed by the ``interplay" between the collective liquid drop nature and single-particle levels near to the Fermi surface \cite{Bohr_Mottelson_vol1, Bohr_Mottelson_vol2}. In general, the beta-stable nuclei close to the major shell closures (like 8, 20, 28, 50, 82, and 126) exhibit nearly spherical ground state shapes. On moving away from these shell closures, the evolution of shell structure of valence nucleons primarily governs the nuclear shapes \cite{casten1990nuclear} such as in the rare-earth region \cite{rareearth1, ReviewByVerney2025}. However, the nuclear shapes do not always evolve smoothly with valence nucleons and complexity of nuclear forces often makes the predictions of nuclear shapes a tricky question.

Nuclei in the sd-shell region populate the region between two closed shells from the magic numbers 8 to 20 and constitute one of the simplest sets of nuclei tractable by the nuclear microscopic theories. This region serves as an ideal testing ground for the nuclear shell-model and the underlying nuclear interactions \cite {DaoPRC2022, WernerNPA1996, BenderPRC2008}. Furthermore, it provides a favorable scenario for investigating the shape coexistence where different shapes manifest in separate but closely lying excited states  
\cite{GARRETT_PPNP2022, OtsukaPPNP2024, Heyde2011, Taniguchi2009_sdNuclei, APRAHAMIAN2025,Bonatsos2023,Li_JPG2016}, and also for shape fluctuations where a nuclear state itself lacks a sharply defined intrinsic deformation and favors a range of shapes due to soft energy surface \cite{cea, Poves2020}. Several calculations have been performed in the $sd$-shell region using the sophisticated modern microscopic theories, \cite {DaoPRC2022, WernerNPA1996, BenderPRC2008, Taniguchi2009_sdNuclei, MyaingHagino2008, MeiHaginoPRC2018, TensorLiSagwa2013}, revealing diverse predictions for nuclear shapes. For example, the relativistic mean-field calculations predict an oblate shape for the ground state of $^{30}$Si \cite{MyaingHagino2008}, the nucleus of interest in this study. However, when beyond mean-field approximations are included, the resulting Potential Energy Surface (PES) becomes broad and centered around spherical shape \cite{MeiHaginoPRC2018}. Interestingly, Skyrme-HF calculations with inclusion of one set of tensor interaction (T$_{64}$ and T$_{66}$) exhibit two clear minima in the PES for $^{30}$Si, one corresponding to a prolate shape and the other to an oblate shape \cite{TensorLiSagwa2013}. In contrast, the inclusion of another set of tensor interaction (T$_{22}$ and T$_{44}$) leads to a very broad and flat minimum around spherical shape \cite{TensorLiSagwa2013}. Notably, this isotope has also received attention for the exploration of sub-shell gap at N=16 \cite{Obertelli2005_NewMagicNum, PELTIER2025139576}.   

On the experimental front, several measurements using hadron scattering such as neutrons \cite{Leo_neutron_28Si}, protons \cite{Fabrici1980, Cole1966}, deuterons \cite{Niewodniczanski1964, BERG1972211},  tritons \cite{PEARCE_Triton30Si}, $\alpha$ particles \cite{KokameAlpha, Siemaszko_JPG1994} and electromagnetic probes--electron scattering \cite{Horikawa1971} and Coulomb excitation \cite{FEWELL1_PRL_979, BALLNPA1980} have been carried out in the $sd$-shell region. In particular, in the case of $^{30}$Si, quite contrasting results on the ground state shape have been obtained. For example, $\alpha $-scattering \cite{Siemaszko_JPG1994} and deuteron scattering \cite{BERG1972211} suggest it to be a prolate shaped nucleus, whereas, measurements using Coulomb excitation have shown it to be spherical as well as oblate shaped nucleus \cite{FEWELL1_PRL_979, Pritychenko:2013gwa,SCHWALM1977425}. These conflicting results from both experimental as well as theoretical fronts highlight the need for further investigations to identify the ground-state structure of $^{30}$Si. 
\begin{figure}[t]
\centering\includegraphics[trim= 0.2mm 0.2mm 0.2mm -2mm, clip, height=0.13\textheight]{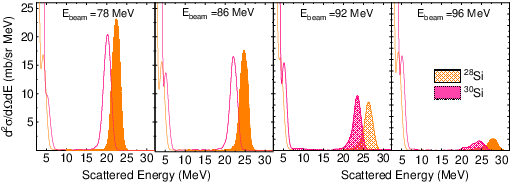}
\caption{\label{28-30SiQELPlots} Typical backward-angle (160$^{\circ}$) energy spectra for $^{28}$Si and $^{30}$Si at four beam energies. The filled curves above 15~MeV mark the QEL events, while those below 10~MeV arise mainly from evaporated light charged particles (see text).
}
\end{figure}

\begin{figure}[t]
\centering\includegraphics[trim= 0.2mm 0.2mm 0.2mm -2mm, clip, height=0.5\textheight]{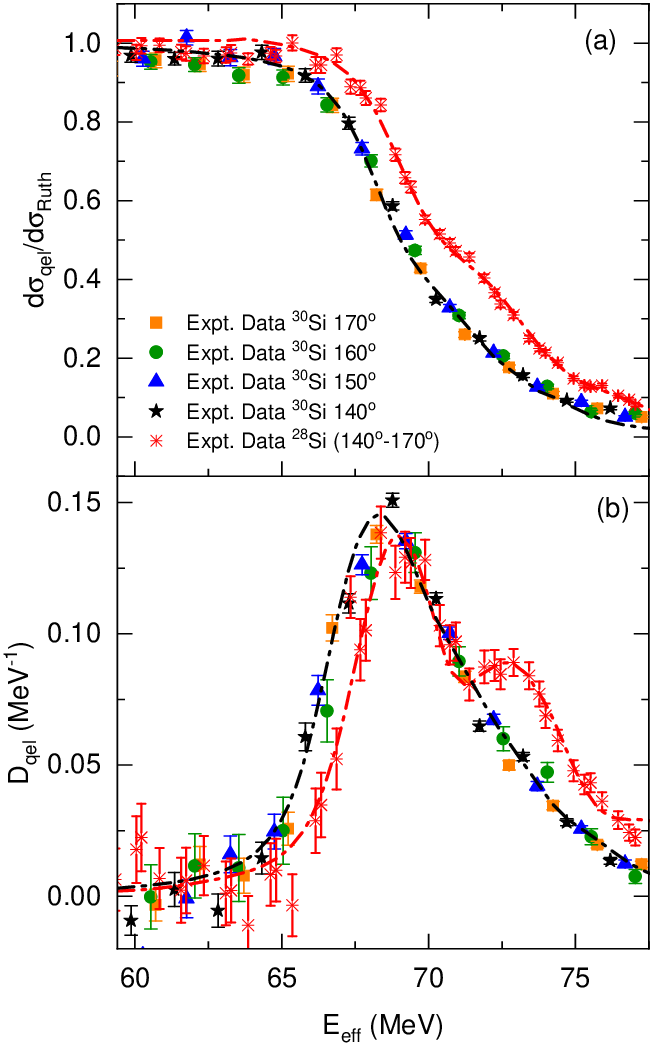}
\caption{\label{30Si_Expt_BD} Quasi-elastic excitation function (panel (a)) and determined quasi-elastic barrier distribution (panel (b)) at four backward angles for $^{28, 30}$Si + $^{90}$Zr reactions. The dash-dotted lines are shown to guide the eye.}
\end{figure}

Discrepancies about the ground state shapes inferred from high energy inelastic scattering and Coulomb excitations have also been reported previously in the other sd-shell nuclei such as $^{24}$Mg  and $^{28}$Si, as discussed in Refs. \cite{ykgplb2020, ykgplb2023}. Whereas, we have established using quasi-elastic (QEL) scattering in the framework of coupled-channels (CC) analyses that $^{24}$Mg and $^{28}$Si posses uniquely prolate and oblate shaped in their respective ground states \cite{ykgplb2020, ykgplb2023, ykgprc2025}. The strong sensitivity of QEL scattering with ground state deformation has also been employed in the rare-earth region \cite{Jia2014}. Both, inelastic and QEL scattering involve transition
strengths which are closely related to the ground state density distribution. However, QEL scattering offers enhanced sensitivity by exploiting the energy derivative of the QEL excitation function. In contrast, high-energy hadron and electron scattering tend to average over instantaneous shape fluctuations \cite{Glauber1959}.

We have carried out experimental investigations using QEL scattering of $^{28}$Si and $^{30}$Si projectiles off the $^{90}$Zr target, analyzed within the framework of coupled-channels (CC) calculations. These results are further complemented by full-space shell-model analyses. The results for the $^{28}$Si+$^{90}$Zr system have been reported earlier in Ref. \cite{ykgplb2023}. In this Letter, we report a comprehensive study combining both experimental and theoretical investigations to probe the ground state shape and structure of $^{30}$Si. Importantly, our findings suggest the presence of shape fluctuations in the ground state of $^{30}$Si, exhibiting a behavior that is markedly different from its neighboring isotopes, $^{28}$Si and $^{32}$Si.

Quasi-elastic measurements were carried out using $^{28, 30}$Si DC beams from BARC-TIFR 14 MV Pelletron accelerator facility.  A spherical closed-shell target combined with high charge product of projectile and target ($\mathrm{Z_{P}Z_{T}}$) is an appropriate choice 
because the channel-coupling form factor in quasi-elastic scattering 
is proportional to $\mathrm{Z_{P}Z_{T}}$, significantly enhancing sensitivity to the ground-state structure of projectile \cite{HaginoPTP2012}. In the present experiment, considering the available beam energies of $^{30}$Si, we have chosen $^{90}$Zr as a target with $\mathrm{Z_{P}Z_{T}}$=560. Experimental techniques used for the QEL measurements have been presented earlier in Ref. \cite{ykgplb2023} along with the results from QEL scattering of $^{28}$Si beam off $^{90}$Zr target. The gross features of the experimental setup are provided here again.  Highly enriched ($>$95\%) $^{90}$Zr (150 $\mu$g/cm$^{2}$) deposited in oxide form on $^{12}$C (40 $\mu$g/cm$^{2}$) backing was used as the target.  Quasi-elastic events were detected using four very thin ($\simeq$ 15 $\mu m$) silicon surface barrier (SSB) detectors placed at 140$^{\circ}$, 150$^{\circ}$, 160$^{\circ}$, and 170$^{\circ}$ with respect to the beam direction. The angular opening of each detector was restricted to close to $\pm 1^{\circ }$. Two more SSB detectors (1 mm), were mounted at 20$^{\circ}$ in the reaction plane on either side of the beam direction for Rutherford normalization and beam monitoring purposes. Each of these monitor detectors was placed at a distance of 48 cm from the target with a collimator of 1 mm diameter. Rutherford scattering peak of $^{90}$Zr was well separated from that of the  $^{12}$C (backing) and $^{16}$O (present in ZrO$_{2}$ target) at forward angles ($\pm$20$^{\circ}$).

Beam energies used were in the range of 70 to 102 MeV in steps of 2-MeV. At every change of beam-energy, the transmission of the beam was maximized through a collimator of 5 mm diameter, enabling a halo-free beam. The solid-angle ratios of monitor to back-angle thin detectors were experimentally determined from  Rutherford scattering of $^{28}$Si projectile off $^{197}$Au (150 $\mu$g/cm$^{2}$) target at 70 and 72 MeV beam energies.  The quasi-elastic events consist of Projectile Like Fragments (PLFs), including elastic and inelastic scattering of projectile off the target,  projectile and/or target excitations, and transfer events.  At the backward angles in the present mass and energy region, the reaction products include PLFs and evaporated particles. The PLFs were clearly distinguished from evaporated Light Charged Particles (LCPs) using pulse height analysis. Typical energy spectra of backward angle scatterings for $^{28}$Si and $^{30}$Si are shown in the Fig. \ref{28-30SiQELPlots} at a laboratory angle of 160$^{\circ}$ for four typical beam energies. In each panel of the Fig. \ref{28-30SiQELPlots}, the filled curves beyond 15 MeV scattered energy, represent the QEL events. The events below 10 MeV are mostly due to the evaporated LCPs.
\begin{figure}[t]
\centering\includegraphics[trim= 0.2mm 0.2mm 0.2mm -2mm, clip, height=0.3\textheight]{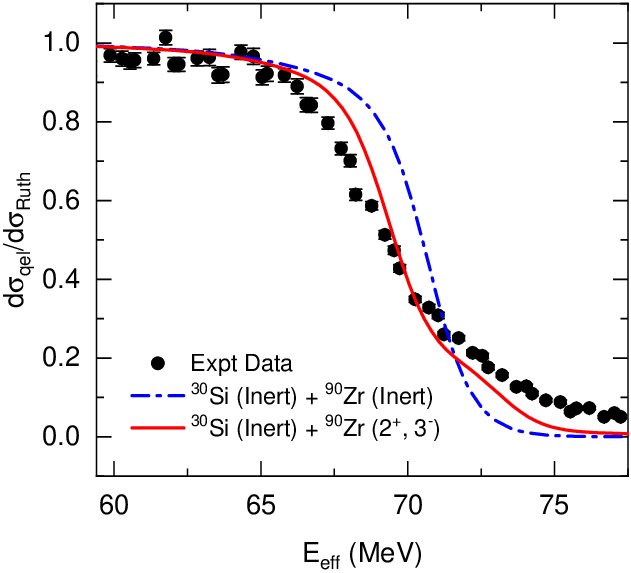}
\caption{\label{30Si_Inert_vib} Quasi-elastic excitation function for $^{30}$Si + $^{90}$Zr reaction. The dash-dotted (blue) and solid lines (red) show CC calculations for the two respective cases: (i) both $^{30}$Si and $^{90}$Zr nuclei are inert and (ii) $^{30}$Si is an inert nucleus and  vibrational couplings of $^{90}$Zr including (2$^{+}$) state at 2.18 MeV and the octupole (3$^{-}$) state at 2.75 MeV.}
\end{figure}

Differential cross section for quasi-elastic events at each beam energy was normalized with Rutherford scattering cross section. The center-of-mass energy ($E_\mathrm{c.m.}$) was corrected for centrifugal effects at each angle as follows \cite{Timmers1995, HaginoPRC2004, BKN2007, Piasecki2005}:
\begin{equation}
E_\mathrm{eff} =\frac{2E_\mathrm{c.m.}}{(1+\mathrm{cosec}(\theta_\mathrm{c.m.}/2))},
\end{equation}
where $\theta_\mathrm{c.m.}$ is the center-of-mass angle.  The quasi-elastic excitation function for the $^{30}$Si + $^{90}$Zr reaction is shown in the Fig. \ref{30Si_Expt_BD}(a) at the four backward angles. It is seen that the quasi-elastic excitation functions at different laboratory angles join quite smoothly. Quasi-elastic barrier distribution $D_\mathrm{qel}$ ($E_\mathrm{eff}$) from quasi-elastic excitation function was determined using the relation \cite{Timmers1995}:
\begin{equation}
D_{qel}(E_\mathrm{eff}) = -\frac{d}{dE_\mathrm{eff}}\bigg[ \frac{d\sigma_\mathrm{qel}(E_\mathrm{eff})}{d\sigma_\mathrm{R}(E_\mathrm{eff})}\bigg],
\end{equation}
where $d\sigma_\mathrm{qel}$ and $d\sigma_\mathrm{R}$ are the differential cross sections for the quasi-elastic and Rutherford scatterings, respectively. A point difference formula is used to evaluate the barrier distribution, with the energy step of 3.1 MeV in the laboratory frame of reference. This barrier distribution determined from backward angle QEL scattering provides a good representation of Fusion Barrier Distribution (FBD) \cite{ykgplb2020, ykgplb2023, ykgprc2025}. Similar to the excitation function, the barrier distribution determined from excitation functions at different laboratory angles joins quite smoothly as shown in the Fig. \ref{30Si_Expt_BD}(b). The smooth joining of the data in excitation function as well as derived barrier distribution, ensure correct identification of the quasi-elastic scattering events.

\begin{figure}[t]
\centering\includegraphics[trim= 0.2mm 0.2mm 0.2mm -2mm, clip, height=0.3\textheight]{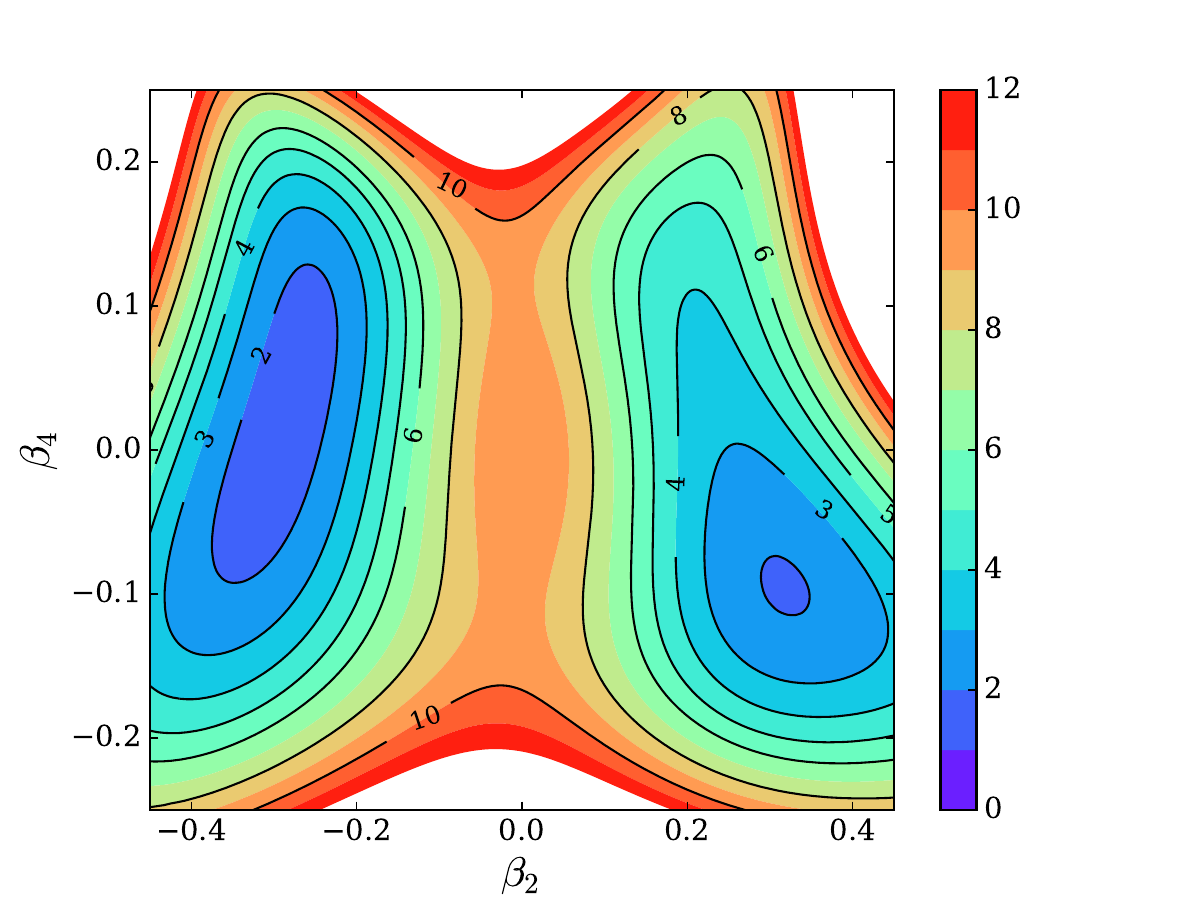}
\caption{\label{Chisq} A reduced $\chi^{2}$ distribution in the two dimensional space of $\beta_{2}$ and $\beta_{4}$ of $^{30}$Si, determined by comparing experimental QEL excitation function for $^{30}$Si+$^{90}$Zr scattering with CC calculations (see text).}
\end{figure}


QEL excitation function and FBD for $^{30}$Si + $^{90}$Zr reaction are also compared with that for $^{28}$Si + $^{90}$Zr reaction (reported in Ref. \cite{ykgplb2023}) in Fig. \ref{30Si_Expt_BD} (a) and (b), respectively, where dash-dotted lines are shown to guide the eye. It is seen that the shapes of excitation functions and FBDs are significantly different
for these two projectiles, differing by just two additional neutrons in $^{30}$Si. In the case of $^{28}$Si, a prominent shoulder like structure is seen in the FBD, making it an asymmetric distribution, whereas it is quite smooth  and symmetric distribution for $^{30}$Si. Moreover, entire FBD for $^{28}$Si is observed to be shifted to higher barrier heights (by 2-3 MeV) relative to that of $^{30}$Si. It is quite remarkable that the data for both $^{28}$Si and $^{30}$Si were taken using the same experimental setup at the same time and data analysis is performed using the same methodology, still the observations are altogether in sharp contrast to each other. These pronounced differences point to fundamentally different ground state structure of $^{30}$Si from that of $^{28}$Si \cite{ykgplb2023}.
\begin{figure}[t]
\includegraphics[trim= -0.2mm 0.2mm 5mm -0.2mm, clip, height=0.24\textwidth]{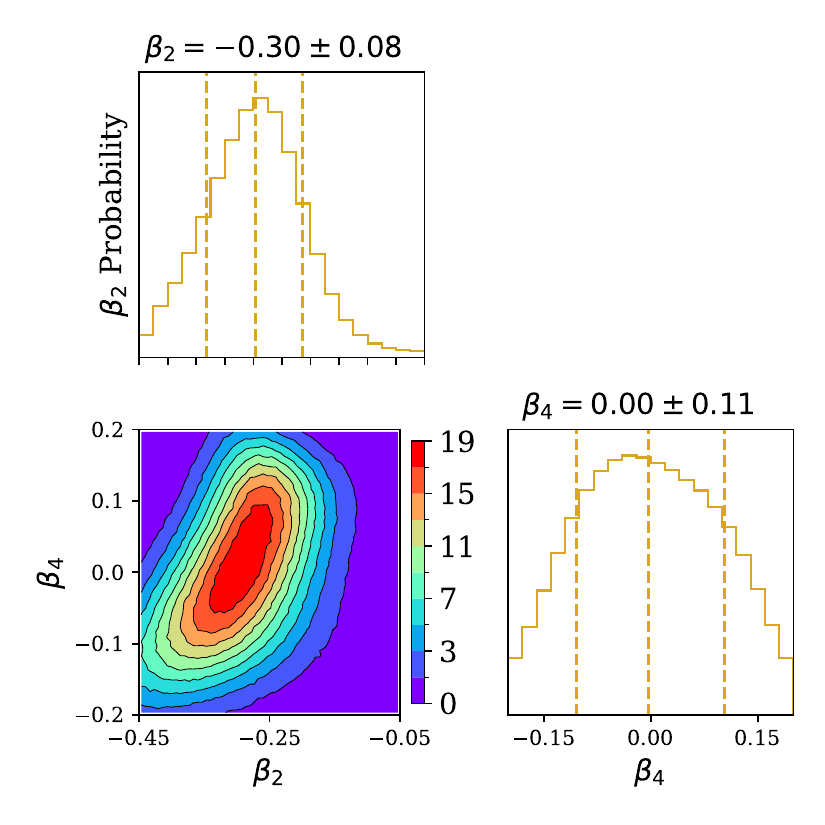}
\includegraphics[trim= 11mm 0.2mm 0.1mm -0.2mm, clip, height=0.24\textwidth]{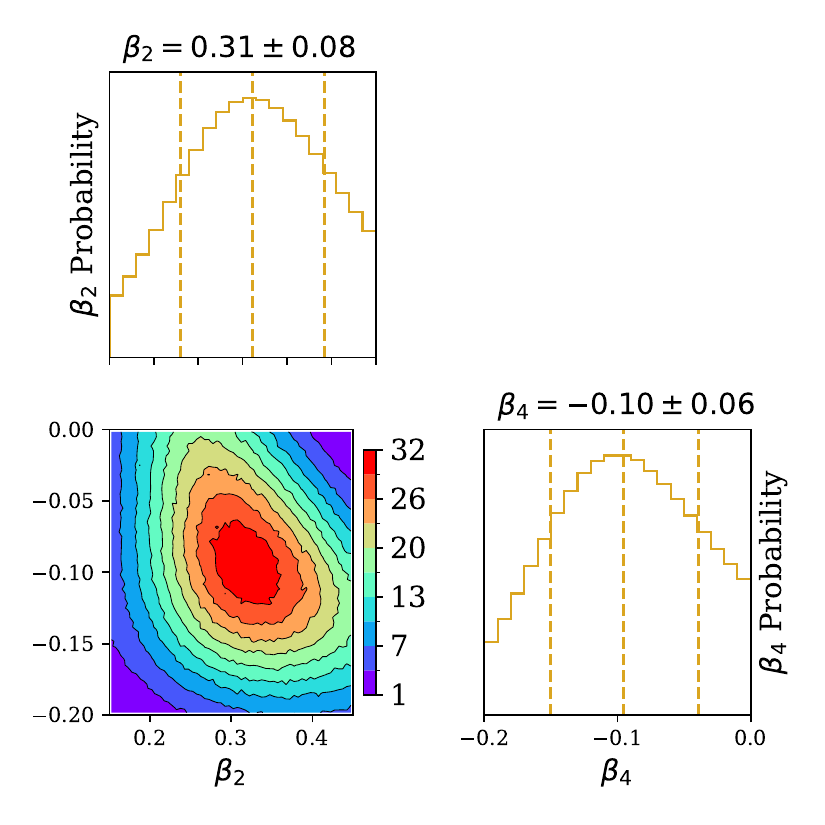}
\caption{\label{probability_dists}  Two-dimensional probability distributions for $\beta_2$ and $\beta_4$ of for $^{30}$Si, resulting from the MCMC simulations \cite{ykgplb2020}. The left and right side simulations correspond to oblate and prolate minima, respectively (see Fig. \ref{Chisq}). }
\end{figure}

In order to understand the ground state structure of $^{30}$Si, Coupled-Channels (CC) calculations were performed using a modified version of CCFULL code \cite{ccqel} for quasi-elastic scattering. Wood-Saxon type optical model potentials were used for both real as well as imaginary parts. The optical model parameters (OMPs) for the real potential were grossly estimated from Broglia-Winther potentials \cite{HaginoPTP2012}, and those were further refined so that coupled calculation can reproduce the experimental data as best as possible.

The CC calculations were carried out first without including any channel coupling (inert).
These  uncoupled calculations failed to reproduce the experimental data as shown in the Fig. \ref{30Si_Inert_vib}. The dominance of vibrational couplings of $^{90}$Zr in QEL scattering as well as in fusion processes has been demonstrated previously for a variety of projectiles \cite{ykgplb2020, kalkal2005, ykgplb2023, ykgprc2025}. In view of this, the CC calculations were further performed  including the vibrational couplings of the  target, $^{90}$Zr. The quadrupole 2$^{+}$ state at 2.18 MeV and the octupole 3$^{-}$ state at 2.75 MeV were considered, with coupling strengths of 0.089 and 0.211, respectively \cite{ykgplb2020, kalkal2005}. However, as evident from Fig.~\ref{30Si_Inert_vib}, even after including the vibrational couplings of $^{90}$Zr while treating $^{30}$Si as inert, the CC calculations still deviate significantly from the experimental data. This highlights the necessity of incorporating additional degrees of freedom of the projectile within the CCFULL framework.

First, we look for the deformed ground-state features of $^{30}$Si using CC calculations. These were performed by including rotational couplings of $^{30}$Si and vibration couplings of  $^{90}$Zr. Rather than using rigid-rotor model, experimental values for the excitation energies of the first 2$^{+}$ (2.235 MeV) and the 4$^{+}$ (5.279 MeV) states of $^{30}$Si were used. Along with quadrupole deformation, the ground-state hexadecapole  deformation of $^{30}$Si was also included in the CC calculations. In order to reproduce the experimental data on quasi-elastic excitation function and the barrier distribution, the CC calculations were carried out in a large parameter space of quadrupole ($\beta_{2}$) and hexadecapole ($\beta_{4}$) deformations. The calculations were performed over a grid of $\beta_{2}$ [from -0.50 (oblate) to +0.50 (prolate)] and $\beta_{4}$ [from -0.40 to +0.40] including 
$\beta_{2}=0$ and $\beta_{4}=0$, and with a step size of 0.01 for both the parameters. Coulomb and nuclear parts were kept at the same values for both quadrupole and hexadecapole deformations. First three rotational states of $^{30}$Si, namely,  0$^{+}$, 2$^{+}$, and 4$^{+}$, were included in the CCFULL calculations. The coupling to the 6$^{+}$ state was confirmed to give a negligible contribution  to the QEL excitation function. A large number of CC calculations were performed using the ``ANUPAM" supercomputer of BARC.

The $\chi^{2}$ was calculated between the experimental QEL scattering excitation function and CC calculations for each combination of $\beta_{2}$ and $\beta_{4}$ using following equation;
\begin{equation}
\chi^{2}(\beta_{2},\beta_{4})=\sum_{i=1}^{N}\frac{[Y_{i}-f(\beta_{2},\beta_{4})]^2}{\sigma_{i}^{2}},
\label{Chi}
\end{equation}
where $Y_{i}$ represents the experimental value of the excitation function at the $i^{th}$ energy point, $\sigma_{i}$ is the uncertainty in the data, and $f(\beta_{2},\beta_{4})$ represents corresponding CCFULL calculation for a particular combination of $\beta_{2}$ and $\beta_{4}$. In  Eq. \ref{Chi}, summation runs over all the data points ($N$) in the effective energy, $E_\mathrm{eff}$ range of 63.8 to 72.7 MeV.  A reduced $\chi^{2}$ is computed as $\chi^{2}_\mathrm{red}$=$\chi^{2}(\beta_{2},\beta_{4})$/DOF, where DOF refers to degrees of freedom. The DOF value is 76 in the present data set for $^{30}$Si+$^{90}$Zr QEL excitation function. The $\chi^{2}_\mathrm{red}$-distribution thus obtained in the two-dimensional space of $\beta_{2}$ and $\beta_{4}$ is shown in Fig. \ref{Chisq}. It shows two minima,  one corresponding to oblate (left contour) and another for prolate shape (right contour).  Both the minima are observed to be of similar depths with $\chi^{2}_\mathrm{red}\approx 1.9$ as can be seen from the Fig. \ref{Chisq}.

In order to obtain the quantitative values of $\beta_{2}$ and $\beta_{4}$ for $^{30}$Si and their associated uncertainties corresponding to prolate and oblate shapes from present data of quasi-elastic excitation function, Bayesian analyses with a Markov-Chain Monte Carlo (MCMC) framework were carried out \cite{ykgplb2020}. Probability distributions obtained from the Bayesian analyses corresponding to both the minima in the $\chi^{2}_\mathrm{red}$-distribution are shown in Fig. \ref{probability_dists}. The best-fitted values of $\beta_{2}$ and $\beta_{4}$ parameters corresponding to $\chi^{2}_\mathrm{red}$$\approx$1.9 are -0.30$\pm$0.08 and 0.00$\pm$0.11, respectively corresponding to oblate minimum and  +0.31$\pm$0.08 and -0.10$\pm$0.06, respectively corresponding to prolate minimum. 

\begin{figure}[t]
\centering\includegraphics[trim= 0.2mm 0.2mm 0.2mm -0.2mm, clip, height=0.66\textwidth]{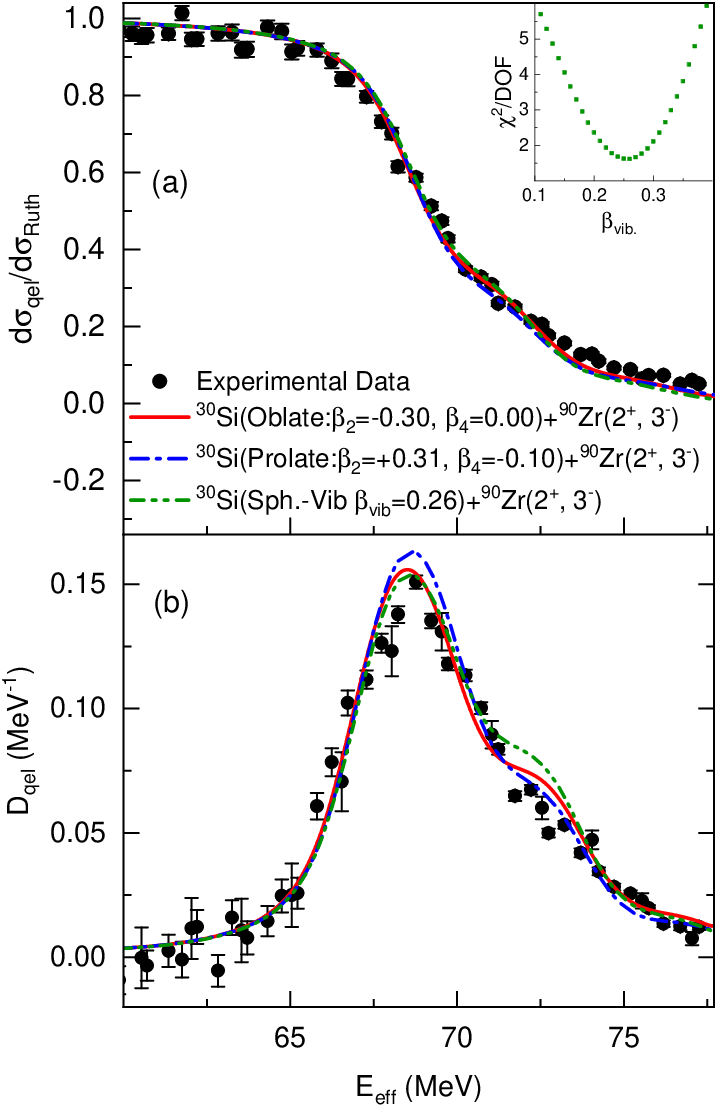}
\caption{\label{30Si_Def_vib}  Quasi-elastic excitation function (a) and barrier distribution (b) for 
$^{30}$Si + $^{90}$Zr system. The solid (red), dash-dotted (blue), and 
dash-dotted (green) curves show CC calculations for oblate, prolate, 
and spherical shapes of $^{30}$Si, respectively. The inset in (a) presents 
$\chi^{2}$/DOF as a function of the vibrational coupling strength, 
$\beta_{\mathrm{vib}}$.
 }
\end{figure}

The coupled-channels calculations based on either parameter set reproduce the 
quasi-elastic scattering data within the experimental uncertainties as shown in the Fig. \ref{30Si_Def_vib}. Thus, the $\chi^{2}$ landscape reveals two nearly degenerate minima—corresponding to oblate and prolate configurations—indicating that the present data cannot uniquely determine the ground-state deformation of $^{30}$Si. This outcome contrasts markedly 
with the $^{28}$Si, for which identical measurements and analysis procedures yield a clear and isolated oblate minimum.

The high value of $\chi^{2}_\mathrm{red}$ at $\beta_{2}=0$ in Fig. \ref{Chisq} just reflects the inadequacy of the rotational coupling matrix for a spherical case ($\beta_{2}=0$) \cite{HaginoPTP2012}. In order to investigate the explicitly spherical ground-state shape of $^{30}$Si, the CC calculations were performed with the vibrational coupling matrix for $^{30}$Si. These calculations were carried out over a wide range of vibrational coupling strengths ($\beta_\mathrm{vib.}$) of $^{30}$Si from 0 to 0.5 in a step size of 0.01. Using the Eq. \ref{Chi}, the $\chi^{2}$/DOF was obtained as a function of $\beta_{vib.}$ in the effective energy range of 63.8 to 72.7 MeV. It is shown as an inset in Fig. \ref{30Si_Def_vib}(a). The best fit $\beta_\mathrm{vib.}$ was obtained to be 0.26 $\pm$0.01 corresponding to $\chi^{2}_\mathrm{red}\approx 1.7$. The CC calculations with  $\beta_\mathrm{vib.}$=0.26 are also included in Fig. \ref{30Si_Def_vib}. 

It is observed that CC calculations based on this spherical-vibrational 
description reproduce the experimental data with a quality comparable to the 
prolate and oblate rotational descriptions. Thus, the 
$\chi^{2}$ surfaces (Figs.~\ref{Chisq} and \ref{30Si_Def_vib}) exhibit three well-defined minima, corresponding to oblate, prolate, and spherical shapes with $\chi^{2}_\mathrm{red}$ values lying in the range of 1.7 to 1.9. Consequently, all the three distinct configurations yield a similar good agreement with the present data for $^{30}$Si.

The adopted value of reduced electric quadrupole transition probability for the $0_1^+ \to 2_1^+$ transition in $^{30}$Si is $0.0215(10)~e^{2}\mathrm{b}^{2}$~\cite{BE2}. 
Using the standard relation, $\beta = (4\pi /3Z R_0^2) [B(\mathrm{E}2)\uparrow/e^2]^{1/2}$, this corresponds to a deformation parameter $\beta = 0.315(7)$~\cite{BE2}. It is well known that the simple linear relation between $\beta$ and $B(\mathrm{E}2)$ neglects higher-order terms in the quadrupole shape invariants \cite{BE2-beta2-beta4}. Retaining only the leading term systematically overestimates $\beta$ for a given $B(\mathrm{E}2)$, with the discrepancy increasing for larger deformations, as clearly seen in the case of $^{24}$Mg~\cite{ykgplb2020}. In view of this, the deformation parameters determined in the present work for 
all three configurations fall within uncertainties consistent with the currently adopted
$B(E2)$ value, reinforcing the coexistence of competing shallow minima in the 
energy surface. Taken together, these observations suggest that $^{30}$Si does 
not possess a rigid intrinsic shape; rather, it is a promising candidate for 
large-amplitude shape fluctuations in its ground state.

To explore the microscopic origin of the ground-state shape fluctuations in $^{30}$Si, we examine the low-lying excitation systematics of the even–even $^{28-32}$Si isotopes. $^{28}$Si lies near the midshell region of the $sd$-shell, where both neutrons and protons predominantly occupy the $1d_{5/2}$ orbital at $Z=N=14$. The successive addition of neutrons from $^{28}$Si to $^{32}$Si modifies the underlying single-particle structure and, consequently, the collective response. These changes are reflected in the experimental signatures of collectivity. The $R_{4/2}=E(4^+_1)/E(2^+_1)$ ratios---2.59, 2.36, and $\approx 3.0$ for $^{28}$Si, $^{30}$Si, and $^{32}$Si, respectively---exhibit a non-monotonic evolution. Likewise, the $2^+_1$ excitation energy increases by 456~keV from $^{28}$Si to $^{30}$Si but decreases by 294~keV in $^{32}$Si. Together, these systematics point to a distinct structural anomaly at $^{30}$Si.

\begin{figure}
\centering\includegraphics[trim= 10mm 10mm 25mm 15mm, clip,width=0.4\textwidth ]{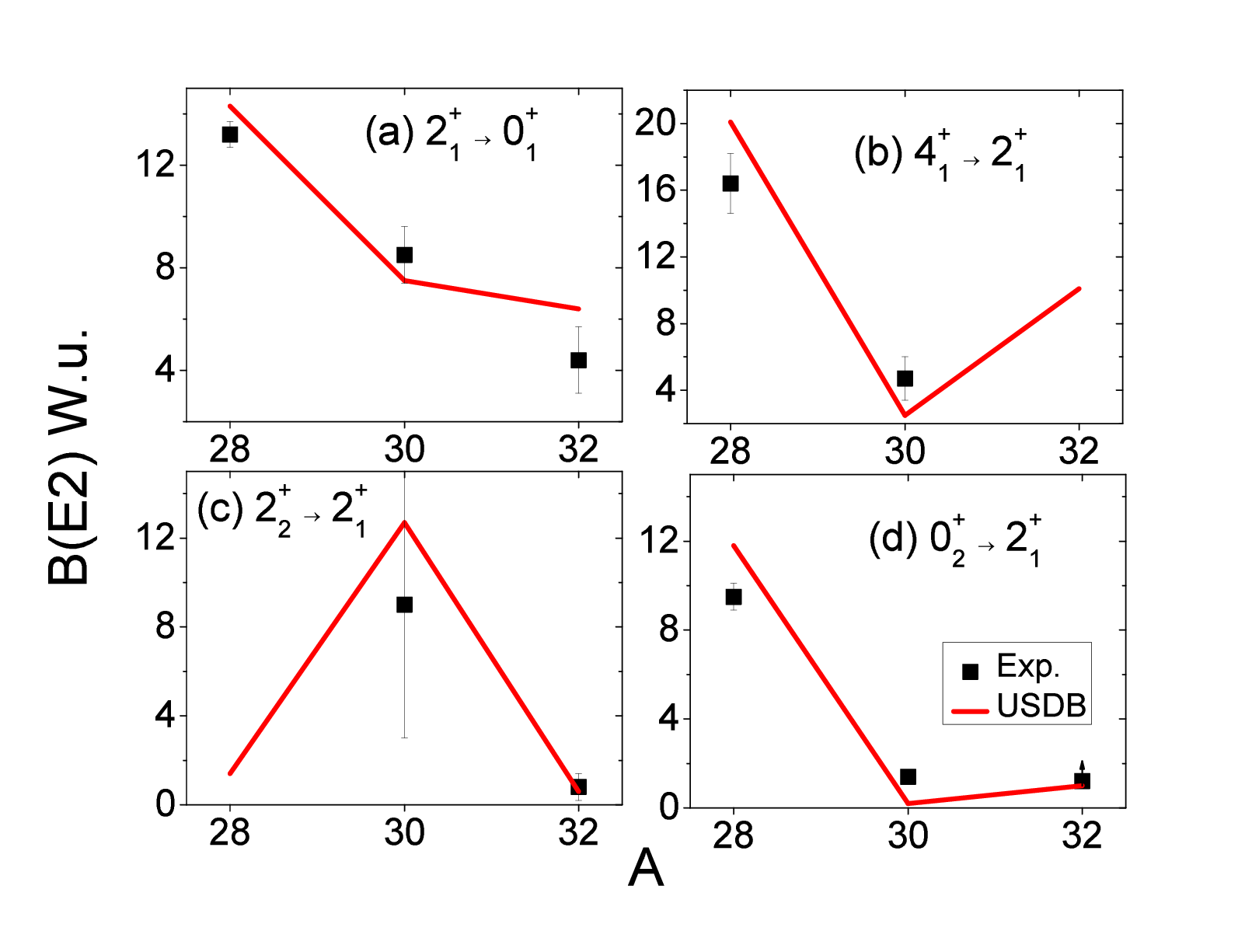}
\caption{\label{fig:SM_be2} Experimental and calculated (with USDB interaction) (a) $B(E2; 2_1^+ \rightarrow 0_1^+)$, (b) $B(E2; 4_1^+ \rightarrow 2_1^+)$, (c) $B(E2; 2_2^+ \rightarrow 2_1^+)$, and (d) $B(E2; 0_2^+ \rightarrow 2_1^+)$, for the even-even $^{28-32}$Si isotopes. In $^{32}$Si, the experimental datum for $B(E2; 0_2^+ \rightarrow 2_1^+)$ represents a lower limit only. }
\end{figure}

Large-scale shell-model calculations employing the USDB interaction~\cite{richter2008} 
(diagonalised with KSHELL~\cite{shimizu2019} and with standard effective charges 
$e_{\pi}=1.35e$, $e_{\nu}=0.35e$) reproduced both level energies of the first- and second-excited low-lying states and $E2$ strengths 
remarkably well for the even–even $^{28-32}$Si isotopes. Striking discontinuities occur at 
$^{30}$Si in going from $^{28}$Si to $^{32}$Si: the $0^+_2$ state drops below 4 MeV, the $2^+_2$ state becomes nearly 
degenerate with $0^+_2$, and the $2^+_1$--$2^+_2$ and $4^+_1$--$4^+_2$ splittings 
change abruptly (see Ref. \cite{supp}). Likewise, except $B(E2;2^+_1\to0^+_1)$ which evolves nearly smoothly, the rest three E2 transitions-$B(E2;2^+_2\to2^+_1)$, $B(E2;4^+_1\to2^+_1)$, 
and $B(E2;0^+_2\to2^+_1)$ also vary sharply at $^{30}$Si as shown in the Fig.~\ref{fig:SM_be2}. Thus, the behavior of low lying states in these neighboring even-even Si isotopes fuels the apprehension about dramatically different low lying structure of $^{30}$Si than $^{28}$Si and $^{32}$Si.

Spectroscopic quadrupole moments of the $2^+_1$ states further support this picture. Shell-model calculations predict $Q_s(2^+_1)=+0.18$, $+0.02$, and $+0.13$\ $eb$ for $^{28}$Si, $^{30}$Si, and $^{32}$Si, respectively. The corresponding measured values are $+0.16(3)$, $-0.05(6)$, and $+0.11(10)$\ $eb$ for $^{28}$Si, $^{30}$Si, and $^{32}$Si, respectively \cite{Stone2016}. The near-zero value of $Q_s$ for $^{30}$Si originates from destructive interference between the calculated proton and neutron $E2$ matrix elements. In $^{28}$Si, where both protons and neutrons predominantly occupy the $d_{5/2}$ orbital, these contributions interfere constructively, producing a sizable quadrupole moment. In contrast, in $^{30}$Si the involvement of active $s_{1/2}$ neutrons, relative to the $d_{5/2}$ protons, induces destructive interference, which may be responsible for the observed shape fluctuations. By $^{32}$Si, the oblate configuration is restored because the neutron occupancy shifts predominantly to the $d_{3/2}$ orbital, re-establishing constructive interference with the proton $d_{5/2}$ contributions and thereby enhancing the quadrupole moment.


Inter-band $E2$ ratios provide additional evidence. Experimentally, $B(E2; 2_2^+ \rightarrow 2_1^+)/B(E2; 2_1^+ \rightarrow 0_1^+)$ turns out to be $\sim 1.1$ and $\sim 0.18$ in $^{30}$Si and $^{32}$Si, respectively. Whereas, it remains unmeasured for $^{28}$Si. Shell-model calculations reproduce the near-SU(3) values of $\sim0.09$ for $^{28}$Si and $^{32}$Si~\cite{dorian2024}, while yielding a much larger value of $\sim1.69$ for $^{30}$Si, consistent with $\gamma$ softness \cite{Cejnar2010}. A second ratio, $B(E2; 0_2^+ \rightarrow 2_1^+) / B(E2; 2_1^+ \rightarrow 0_1^+)$ is nearly zero for $^{30}$Si in both experiment and theory, in contrast to the SU(3)-like value in $^{28}$Si and the moderate value in $^{32}$Si. Nearly zero value of this ratio for $^{30}$Si  conforms to the $\gamma-$soft character \cite{Cejnar2010}.

HFB calculations with the Gogny interaction~\cite{cea,Poves2020} provide further insight into the structural aspects of the Si isotopes. For $^{28}$Si, the calculations predict a sharply defined oblate minimum, in agreement with QEL-scattering results~\cite{ykgplb2023}. In contrast, the deformation-energy surface of $^{30}$Si is notably flat, extending across oblate, prolate, and near-spherical configurations with $\gamma$ values from $0^\circ$ to $60^\circ$, consistent with a pronounced $\gamma$-soft character. The corresponding density profiles~\cite{cea,Poves2020} also show a substantially stronger central density depletion in $^{30}$Si than in $^{28}$Si, hinting at an incipient bubble-like structure possibly linked to enhanced $s_{1/2}$ occupancy~\cite{BubbleNuclei_Perera2022}.

Taken together, the QEL scattering analysis in the framework of CC calculations, HFB–Gogny energy surfaces, density distributions, and shell-model transition strengths consistently indicate that the $^{30}$Si ground state is neither spherical nor well deformed. Instead, it exhibits strong shape fluctuations characteristics of a $\gamma$-soft nucleus. Further theoretical calculations using symmetry insights \cite{Bonatsos2023} and beyond mean field frameworks \cite{Li_JPG2016} would be interesting to probe the ground-state shape of $^{30}$Si. The present experimental results thus provide a useful test-bed for advanced theoretical treatments, against which future models can be evaluated and refined.

In summary, quasi-elastic (QEL) scattering measurements for the 
$^{28,30}$Si+$^{90}$Zr systems were performed at multiple laboratory angles, 
and the resulting QEL excitation functions and fusion barrier distributions 
(FBDs) were extracted. For $^{28}$Si, the FBD exhibits a pronounced 
shoulder-like structure, resulting in a distinctly asymmetric shape, whereas for $^{30}$Si it is 
smooth and symmetric. Coupled-channels (CC) calculations were carried out 
using both deformed-rotational and spherical-vibrational descriptions. The 
rotational analysis included a broad range of quadrupole ($\beta_{2}$) and 
hexadecapole ($\beta_{4}$) deformations, while the vibrational calculations 
covered a wide span of vibrational-coupling strengths. For $^{28}$Si, the CC results 
confirm a uniquely oblate ground state~\cite{ykgplb2023}. In contrast, for 
$^{30}$Si, oblate, prolate, and spherical configurations reproduce the data 
equally well. Shell-model calculations also show no strong spectroscopic 
quadrupole moment for $^{30}$Si, owing to the competing opposite directions favored by valence protons and neutrons, consistent with its soft, fluid-like character. Other 
spectroscopic indicators---including low-lying excitation energies and 
$B(E2)$ ratios---likewise support this $\gamma$-soft behavior. The present study thus indicate that the $^{30}$Si ground state is neither spherical nor well deformed, making it a compelling case of strong shape fluctuations characteristics of a $\gamma$-soft system. It is quite remarkable that the experimental data for both $^{28}$Si and $^{30}$Si were obtained simultaneously using the same setup and analyzed with an identical methodology, yet the sharply differing outcomes point to fundamentally different ground-state structures. Altogether, this study illustrates how the addition of two neutrons beyond the N=Z line in $^{30}$Si alters the delicate nucleonic shape balance when compared with the oblate shaped $^{28}$Si. It would be an interesting future work to extend the coupled-channels approach to take into account the softness of the intrinsic shapes of projectile and/or target.

We are grateful to the operating staff of BARC-TIFR Pelletron staff for the smooth operation of the machine. Y. K. Gupta is thankful to Dr. R. K. Choudhury for discussion at various stages of this work. B. M. gratefully acknowledges the financial support from the HORIZON-MSCA-2023-PF-01 project, ISOON, under Grant No. 101150471. B. M. also thanks P. E. Garrett and P. Van Isacker for useful discussions. U. Garg acknowledges the financial support from National Science Foundation (Grant No. PHY-2310059).


\begin{thebibliography}{10}
\expandafter\ifx\csname url\endcsname\relax
  \def\url#1{\texttt{#1}}\fi
\expandafter\ifx\csname urlprefix\endcsname\relax\def\urlprefix{URL }\fi
\expandafter\ifx\csname href\endcsname\relax
  \def\href#1#2{#2} \def\path#1{#1}\fi

\bibitem{Taniuchi2025}
R.~Taniuchi, Prog. Theor. Expt. Phys. 2025 (2025) ptaf084.

\bibitem{SchatzPR1998}
H.~Schatz, A.~Aprahamian, J.~Görres, M.~Wiescher, T.~Rauscher, J.~Rembges,
  F.-K. Thielemann, B.~Pfeiffer, P.~Möller, K.-L. Kratz, H.~Herndl, B.~Brown,
  H.~Rebel, Phys. Rep. 294 (1998) 167.

\bibitem{AndreyevReport2018}
A.~N. Andreyev, K.~Nishio, \mbox{K-H Schmidt}, Rep. Prog. Phys. 81 (2018)
  016301.

\bibitem{Ibrahim24}
I.~Abdurrahman, M.~Kafker, A.~Bulgac, I.~Stetcu, Phys. Rev. Lett. 132 (2024)
  202501.

\bibitem{NLDBD2017}
J.~Engel, J.~Men\`{e}ndez, Rep. Prog. Phys. 80 (2017) 046301.

\bibitem{Cline_CE_1998}
D.~Cline, Ann. Rev. Nucl. Part. Sci. 36 (1986) 683.

\bibitem{Yang_LASER_2023}
X.~F. Yang, S.~J. Wang, S.~G. Wilkins, R.~F.~G. Ruiz, Prog. Part. Nucl. Phys.
  129 (2023) 104005.

\bibitem{Leo_neutron_28Si}
\mbox{R. De Leo}, G.~D\textquotesingle{}Erasmo, E.~M. Fiore, A.~Pantaleo,
  M.~Pignanelli, Phys. Rev. C 20 (1979) 1244.

\bibitem{Fabrici1980}
E.~Fabrici, S.~Micheletti, M.~Pignanelli, F.~G. Resmini, \mbox{R. De Leo},
  G.~D\textquotesingle{}Erasmo, A.~Pantaleo, Phys. Rev. C 21 (1980) 844.

\bibitem{Cole1966}
R.~K. Cole, C.~N. Waddell, R.~R. Dittman, H.~S. Sandhu, Nucl. Phys. 75 (1966)
  241.

\bibitem{KokameAlpha}
J.~Kokame, K.~Fukunaga, N.~Inoue, H.~Nakamura, Phys. Lett. 8 (1964) 342.

\bibitem{Siemaszko_JPG1994}
M.~Siemaszko, J.~Czakanski, A.~Grzeszczuk, K.~Jankowski, J.~Kisiel,
  B.~Kozlowska, A.~Surowiec, W.~Zipper, A.~Budzanowski, Journal of Physics G:
  Nucl. Part. Phys. 20~(11) (1994) 1789.

\bibitem{238UBeta4PRL}
W.~Ryssens, G.~Giacalone, B.~Schenke, C.~Shen, Phys. Rev. Lett. 130 (2023)
  212302.

\bibitem{nature_shape_2024}
M.~I. \mbox{Abdulhamid et al.,} (STAR~Collaboration), Nature 635 (2024) 67.

\bibitem{RPP_RHIC_shape_2025}
S.~Collaboration, Rep. Prog. Phys. 88 (2025) 108601.

\bibitem{Hagino_RHIC_2025}
K.~Hagino, M.~Kitazawa, Phys. Rev. C 112 (2025) L041901.

\bibitem{Bohr_Mottelson_vol1}
A.~Bohr, B.~R. Mottelson, Nuclear Structure Vol. 1, World Scientific, 1998.

\bibitem{Bohr_Mottelson_vol2}
A.~Bohr, B.~R. Mottelson, Nuclear Structure Vol. 2, World Scientific, 1998.

\bibitem{casten1990nuclear}
R.~F. Casten, Nuclear Structure from a Simple Perspective, Oxford University
  Press, Oxford University Press, 1990.

\bibitem{rareearth1}
Y.~Tsunoda, N.~Shimizu, T.~Otsuka, Phys. Rev. C 108 (2023) L021302.

\bibitem{ReviewByVerney2025}
D.~Verney, Eur. Phys. J. A 61 (2025) 82.

\bibitem{DaoPRC2022}
D.~D. Dao, F.~Nowacki, Phys. Rev. C 105 (2022) 054314.

\bibitem{WernerNPA1996}
T.~R. Werner, J.~A. Sheikh, M.~Misu, W.~Nazarewicz, J.~Rikovska, K.~Heeger,
  A.~S. Umar, M.~R. Strayer, Nucl. Phys. A 597 (1996) 327.

\bibitem{BenderPRC2008}
\mbox{Michael Bender}, \mbox{Paul-Henri Heenen}, Phys. Rev. C 78 (2008) 024309.

\bibitem{GARRETT_PPNP2022}
P.~E. Garrett, M.~Zielińska, E.~Cl\'{e}ment, Prog. Part. Nucl. Phys. 124
  (2022) 103931.

\bibitem{OtsukaPPNP2024}
S.~Leoni, B.~Fornal, A.~Bracco, Y.~Tsunoda, T.~Otsuka, Prog. Part. Nucl. Phys.
  139 (2024) 104119.

\bibitem{Heyde2011}
K.~Heyde, J.~L. Wood, Rev. Mod. Phys. 83 (2011) 1467.

\bibitem{Taniguchi2009_sdNuclei}
Y.~Taniguchi, Y.~Kanada-En'yo, M.~Kimura, Prog. Theor. Phys. 121~(3) (2009)
  533.

\bibitem{APRAHAMIAN2025}
A.~Aprahamian, K.~Lee, S.~R. Lesher, R.~Bijker, Prog. Part. Nucl. Phys. 143
  (2025) 104173.

\bibitem{Bonatsos2023}
D.~Bonatsos, A.~Martinou, S.~K. Peroulis, T.~J. Mertzimekis, N.~Minkov, Atoms
  11 (2023) 117.

\bibitem{Li_JPG2016}
Z.~P. Li, T.~Nikšić, D.~Vretenar, Journal of Physics G: Nuclear and Particle
  Physics 43 (2016) 024005.

\bibitem{cea}
\url{https://www-phynu.cea.fr}.

\bibitem{Poves2020}
A.~Poves, F.~Nowacki, Y.~Alhassid, Phys. Rev. C 101 (2020) 054307.

\bibitem{MyaingHagino2008}
M.~T. Win, K.~Hagino, Phys. Rev. C 78 (2008) 054311.

\bibitem{MeiHaginoPRC2018}
H.~Mei, K.~Hagino, J.~M. Yao, T.~Motoba, Phys. Rev. C 97 (2018) 064318.

\bibitem{TensorLiSagwa2013}
A.~Li, X.~R. Zhou, H.~Sagawa, Prog. Theor. Expt. Phys. 2013 (2013) 063D03.

\bibitem{Obertelli2005_NewMagicNum}
A.~Obertelli, S.~P\'eru, \mbox{J. -P. Delaroche}, A.~Gillibert, M.~Girod,
  H.~Goutte, Phys. Rev. C 71 (2005) 024304.

\bibitem{PELTIER2025139576}
J.~F. Peltier, Z.~Y. Xu, I.~Cox, R.~Grzywacz, R.~S. Lubna, N.~Kitamura,
  S.~Neupane, J.~M. Allmond, J.~Christie, A.~A. Doetsch, P.~Dyszel,
  T.~Gaballah, T.~T. King, K.~Kolos, S.~N. Liddick, M.~Madurga, T.~H. Ogunbeku,
  B.~M. Sherrill, K.~Siegl, Phys Lett. B 866 (2025) 139576.

\bibitem{Niewodniczanski1964}
H.~Niewodnicza\'{n}ski, J.~Nurzy\'{n}ski, A.~Strza{\l}kowski,
  J.~Wilczy\'{n}ski, J.~R. Rook, P.~E. Hodgson, Nucl. Phys. 55 (1964) 386.

\bibitem{BERG1972211}
M.~Berg, A.~Hofmann, K.~Thomas, H.~Rebel, G.~W. Schweimer, Phys. Lett. B 42
  (1972) 211.

\bibitem{PEARCE_Triton30Si}
K.~I. Pearce, N.~M. Clarke, R.~J. Griffiths, P.~J. Simmonds, D.~Barker,
  J.~B.~A. England, M.~C. Mannion, C.~A. Ogilvie, Nucl. Phys. A 467 (1987) 215.

\bibitem{Horikawa1971}
Y.~Horikawa, Y.~Torizuka, A.~Nakada, S.~Mitsunobu, Y.~Kojima, M.~Kimura, Phys.
  Lett. B 36 (1971) 9.

\bibitem{FEWELL1_PRL_979}
M.~P. Fewell, D.~C. Kean, R.~H. Spear, T.~H. Zabel, A.~M. Baxter, S.~Hinds,
  Phys. Rev. Lett. 43 (1979) 1463.

\bibitem{BALLNPA1980}
G.~C. Ball, O.~H\"{a}usser, T.~K. Alexander, W.~G. Davies, J.~S. Forster, I.~V.
  Mitchell, J.~R. Beene, D.~H\"{o}rn, W.~Mclatchie, Nucl. Phys. A 349 (1980)
  271.

\bibitem{Pritychenko:2013gwa}
B.~Pritychenko, M.~Birch, B.~Singh, M.~Horoi, Atom. Data Nucl. Data Tabl. 107
  (2016) 1.

\bibitem{SCHWALM1977425}
D.~Schwalm, E.~K. Warburton, J.~W. Olness, Nucl. Phys. A 293~(3) (1977) 425.

\bibitem{ykgplb2020}
Y.~K. Gupta, B.~K. Nayak, U.~Garg, K.~Hagino, K.~B. Howard, N.~Sensharma,
  M.~{\c S}enyi{\u g}it, W.~P. Tan, P.~D. O'Malley, M.~Smith, R.~Gandhi,
  T.~Anderson, R.~J. deBoer, B.~Frentz, A.~Gyurjinyan, O.~Hall, M.~R. Hall,
  J.~Hu, E.~Lamere, Q.~Liu, A.~Long, W.~Lu, S.~Lyons, K.~Ostdiek, C.~Seymour,
  M.~Skulski, \mbox{B. Vande Kolk}, Phys. Lett. B 806 (2020) 135473.

\bibitem{ykgplb2023}
Y.~K. Gupta, V.~B. Katariya, G.~K. Prajapati, K.~Hagino, D.~Patel, V.~Ranga,
  U.~Garg, L.~S. Danu, A.~Pal, B.~N. Joshi, S.~Dubey, V.~V. Desai, S.~Panwar,
  N.~Kumar, S.~Mukhopadhyay, \mbox{Pawan Singh}, N.~Sirswal, R.~Sariyal,
  I.~Mazumdar, B.~V. John, Phys. Lett. B 845 (2023) 138120.

\bibitem{ykgprc2025}
Y.~K. Gupta, K.~Hagino, D.~Patel, V.~B. Katariya, H.~Vyas, G.~K. Prajapati,
  N.~Sirswal, P.~Singh, B.~N. Joshi, B.~K. Nayak, U.~Garg, Phys. Rev. C 112
  (2025) 034616.

\bibitem{Jia2014}
H.~M. Jia, C.~J. Lin, F.~Yang, X.~X. Xu, H.~Q. Zhang, Z.~H. Liu, Z.~D. Wu,
  L.~Yang, N.~R. Ma, P.~F. Bao, L.~J. Sun, Phys. Rev. C 90 (2014) 031601(R).

\bibitem{Glauber1959}
R.~J. Glauber, High-energy collision theory, in: W.~E. Brittin, L.~G. Dunham
  (Eds.), Lectures in Theoretical Physics, Vol.~1, Interscience, New York,
  1959, pp. 315--414.

\bibitem{HaginoPTP2012}
K.~Hagino, N.~Takigawa, Prog. Theor. Phys. 128 (2012) 1061.

\bibitem{Timmers1995}
H.~Timmers, J.~R. Leigh, M.~Dasgupta, D.~J. Hinde, R.~C. Lemmon, J.~C. Mein,
  C.~R. Morton, J.~O. Newton, N.~Rowley, Nucl. Phys. A 584 (1995) 190.

\bibitem{HaginoPRC2004}
K.~Hagino, N.~Rowley, Phys. Rev. C 69 (2004) 054610.

\bibitem{BKN2007}
B.~K. Nayak, R.~K. Choudhury, A.~Saxena, P.~K. Sahu, R.~G. Thomas, D.~C.
  Biswas, B.~V. John, E.~T. Mirgule, Y.~K. Gupta, M.~Bhike, H.~G. Rajprakash,
  Phys. Rev. C 75 (2007) 054615.

\bibitem{Piasecki2005}
E.~Piasecki, L.~\`{S}widerski, P.~Czosnyka, M.~Kowalczyk, K.~Piasecki,
  M.~Witecki, T.~Czosnyka, J.~Jastrz\k{e}bski, A.~Kordyasz, M.~Kisieli\`{n}ski,
  T.~Krogulski, M.~Mutterer, S.~Khlebnikov, W.~H. Trzaska, K.~Hagino,
  N.~Rowley, Phys. Lett. B 615 (2005) 55.

\bibitem{ccqel}
K.~Hagino, N.~Rowley, A.~Kruppa, Comput. Phys. Commun. 123 (1999) 143.

\bibitem{kalkal2005}
S.~Kalkal, S.~Mandal, N.~Madhavan, E.~Prasad, S.~Verma, A.~Jhingan, R.~Sandal,
  S.~Nath, J.~Gehlot, B.~R. Behera, M.~Saxena, S.~Goyal, D.~Siwal, R.~Garg,
  U.~D. Pramanik, S.~Kumar, T.~Varughese, K.~S. Golda, S.~Muralithar, A.~K.
  Sinha, R.~Singh, Phys. Rev. C 81 (2010) 044610.

\bibitem{BE2}
S.~Raman, C.~W. Nestor, Jr., P.~Tikkanen, At. Data and Nucl. Data Tables 78
  (2001) 1.

\bibitem{BE2-beta2-beta4}
X.-J. Sun, C.-X. Chen, N.~Wang, H.-B. Zhou, Chinese Physics C 42~(12) (2018)
  124105.

\bibitem{richter2008}
W.~A. Richter, S.~Mkhize, B.~A. Brown, Phys. Rev. C 78 (2008) 064302.

\bibitem{shimizu2019}
N.~Shimizu, T.~Mizusaki, Y.~Utsuno, Y.~Tsunoda, Comput. Phys. Commun. 244
  (2019) 372.

\bibitem{supp}
See Supplemental Material.

\bibitem{Stone2016}
N.~J. Stone, Atomic Data and Nuclear Data Tables 111 (2016) 1.

\bibitem{dorian2024}
D.~Frycz, J.~Men\'endez, A.~Rios, B.~Bally, T.~R. Rodr\'{\i}guez, A.~M. Romero,
  Phys. Rev. C 110 (2024) 054326.

\bibitem{Cejnar2010}
P.~Cejnar, J.~Jolie, R.~F. Casten, Rev. Mod. Phys. 82 (2010) 2155.

\bibitem{BubbleNuclei_Perera2022}
U.~C. Perera, A.~V. Afanasjev, Phys. Rev. C 106 (2022) 024321.

\end{thebibliography}

\end{document}


\begin{frontmatter}

\title{Supplementary Material for \text{Quasi-elastic scattering for nuclear ground state structure: An intriguing case of $^{30}$Si}}

\author[BARC,HBNI]{Y. K. ~Gupta\corref{cor1}}
\cortext[cor1]{Corresponding author, Email:ykgupta@barc.gov.in}

\author[GANIL]{B. Maheshwari}
\author[BARC]{G. K. Prajapati}
\author[Amity,IITR]{A. K. Jain}
\author[KYOTO]{K. Hagino}
\author[BARC,HBNI]{B. N. Joshi}
\author[BARC]{A. Pal}
\author[BARC]{N. Sirswal}
\author[BARC,HBNI]{Pawan Singh}
\author[BARC]{S. Dubey}
\author[SIES]{V. V. Desai}
\author[TIFR]{V. Ranga}
\author[BARC]{V. B. Katariya}
\author[SVNIT]{D. Patel}
\author[BARC]{H. Vyas}
\author[IITR]{S. Panwar}
\author[BARC]{B. V. John}
\author[TIFR]{I. Mazumdar}
\author[BARC]{B. K. Nayak}
\author[ND]{U. Garg}

\address[BARC]{Nuclear Physics Division, Bhabha Atomic Research Centre, Mumbai - 400085, India}
\address[HBNI]{Homi Bhabha National Institute, Anushaktinagar, Mumbai 400094, India} 
\address[GANIL]{Grand Acc\'el\'erateur National d'Ions Lourds, CEA/DSM--CNRS/IN2P3, Bvd Henri Becquerel, BP 55027, F-14076 Caen, France}
\address[Amity]{Amity Institute of Nuclear Science \& Technology, Amity University UP, NOIDA}
\address[IITR]{Department of Physics, IIT Roorkee, Roorkee-247667, India}
\address[KYOTO]{Department of Physics, Kyoto University, Kyoto 606-8502, Japan}
\address[SIES]{SIES College of Arts, Science and Commerce, Mumbai-400022, India}
\address[TIFR]{Tata Institute of Fundamental Research, Mumbai 400005, India}
\address[SVNIT]{Department of Physics, Sardar Vallabhbhai National Institute of Technology, Surat -395007, India}
\address[ND]{Department of Physics and Astronomy, University of Notre Dame, Notre Dame, IN 46556, USA}

\date{}
\end{frontmatter}


Large-scale shell-model calculations employing the USDB interaction~\cite{richter2008}, diagonalised with KSHELL~\cite{shimizu2019}, and with standard effective charges 
$e_{\pi}=1.35e$, $e_{\nu}=0.35e$, were carried out. The Shell model USDB interaction is comprised of same proton and neutron $1d_{3/2}$, $1d_{5/2}$, and $2s_{1/2}$ orbitals with the respective single-particle energies of $+2.1117$, $-3.9257$, and $-3.2079$ MeV. The average shell model occupancies of the active orbitals are listed in Table~\ref{tab:sm} along with the averaged orbital angular momentum contributions of protons and neutrons for a given state. With $^{16}$O as a core, the valence six protons predominantly occupy the $1d_{5/2}$ orbital. The contributions from the $1d_{3/2}$ and $2s_{1/2}$ orbitals are also evident in full proton wave-functions which tend to decrease with increasing mass number among the heavier Si isotopes. However, the neutron contributions of these $1d_{3/2}$ and $2s_{1/2}$ orbitals increase with the addition of neutrons, as expected.

\begin{figure*}
\includegraphics[trim= 10mm 5mm 30mm 10mm, clip,width=0.3\textwidth ]{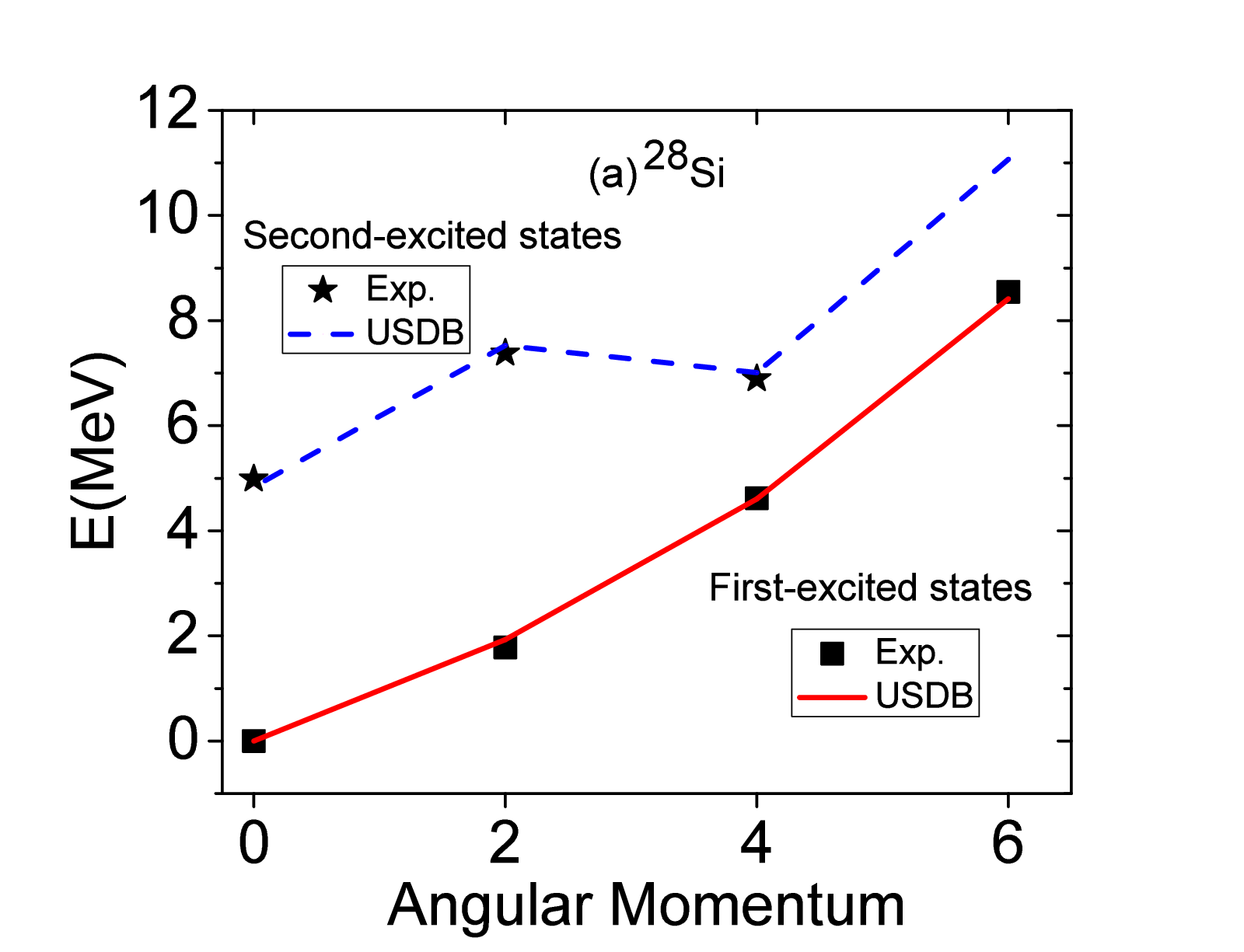}
\includegraphics[trim= 10mm 5mm 30mm 10mm, clip,width=0.3\textwidth ]{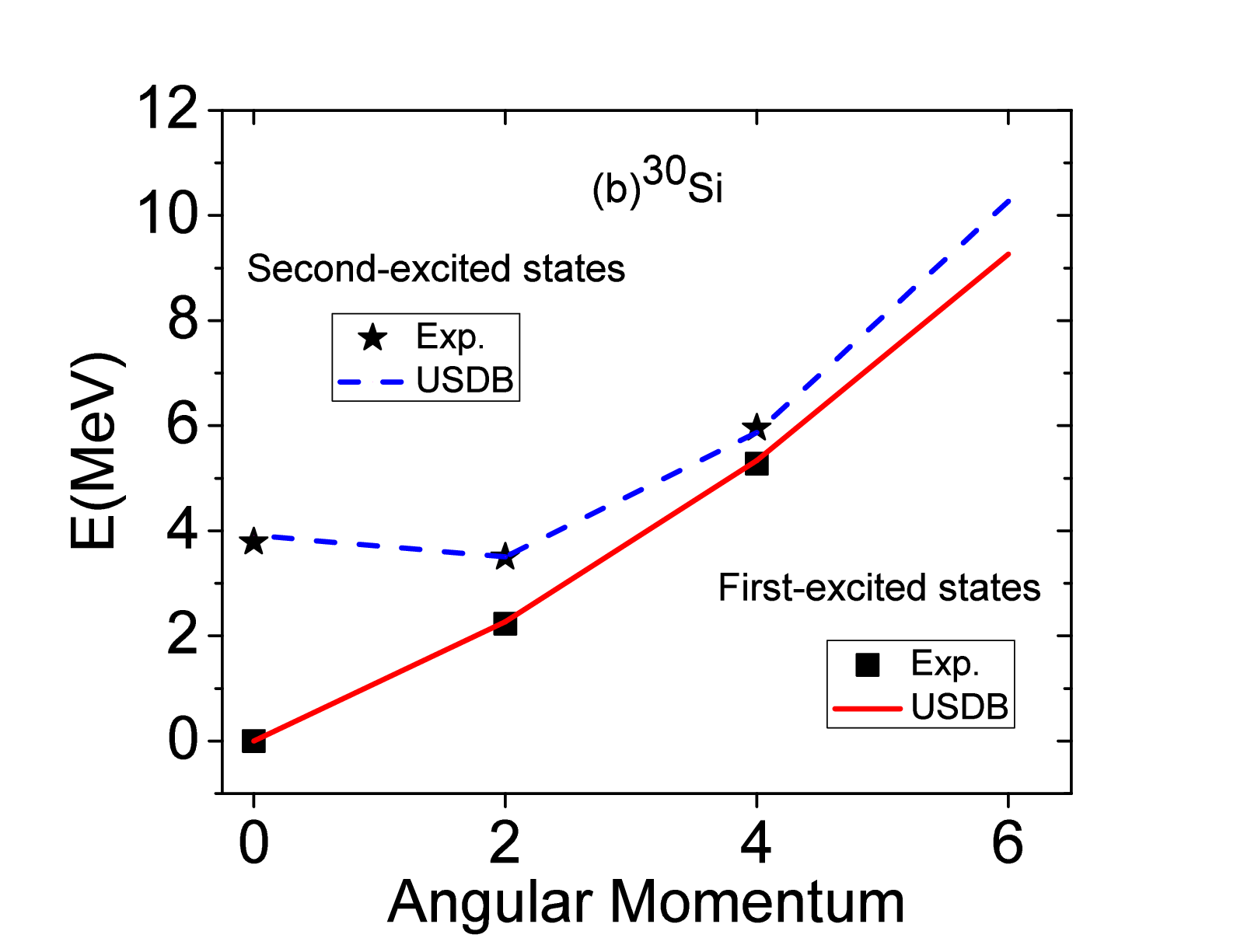}
\includegraphics[trim= 10mm 5mm 30mm 10mm, clip,width=0.3\textwidth ]{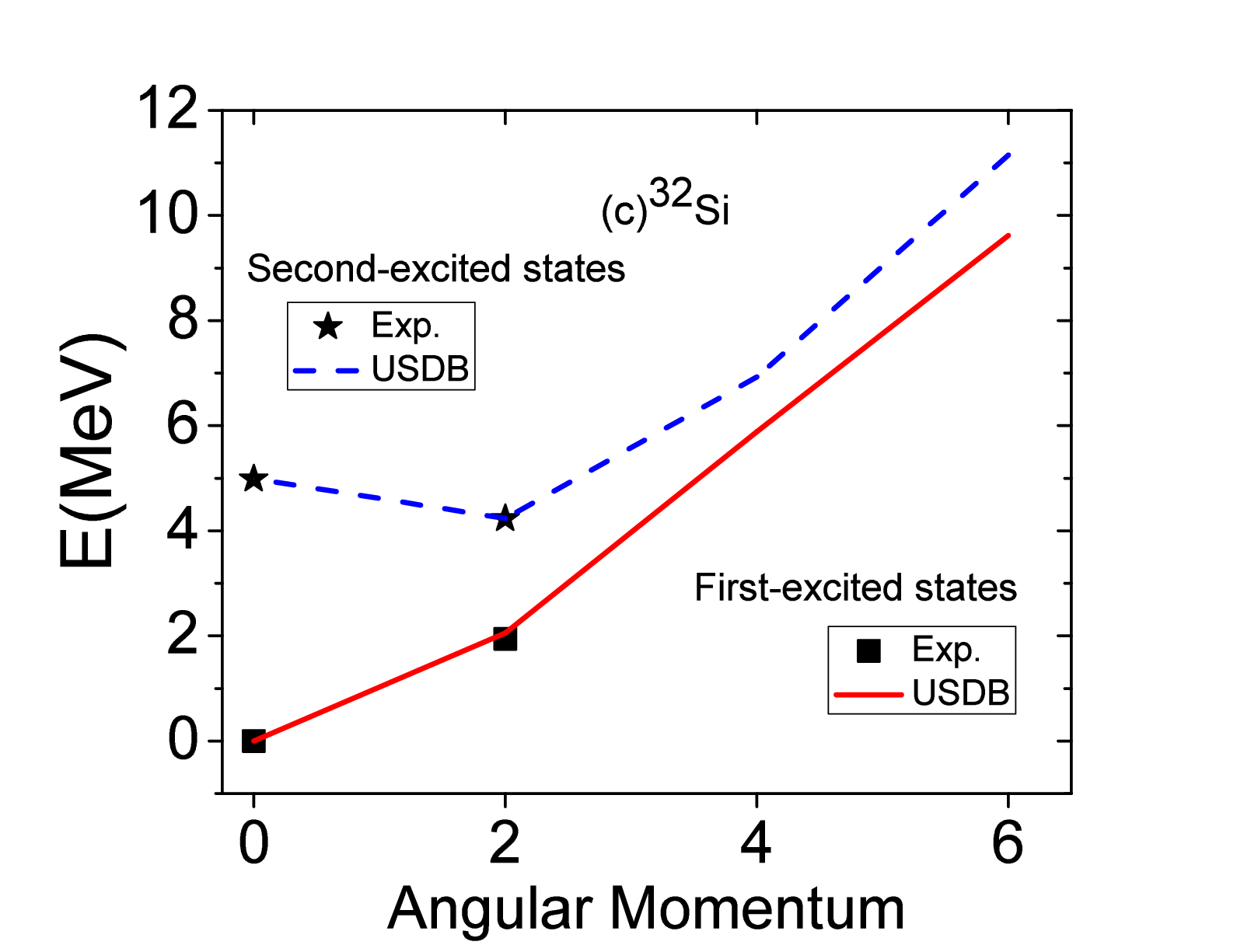}
\caption{\label{fig:SM_levels}  Experimental and calculated (shell model with USDB interaction) energies of the first- and second-excited $0^+$, $2^+$, $4^+$, and $6^+$ states, labeled as angular momentum (in units of $\hbar$) on the x-axis for (a) $^{28}$Si, (b) $^{30}$Si, and (c)$^{32}$Si. }
\end{figure*}

Furthermore, we investigate the quadrupole moments of the second-excited $2_2^+$ states in all three Si isotopes. Currently, no experimental data are available. The shell model predicts spectroscopic quadrupole moments $Q_{s}$ of $-0.070$ \textit{eb},  $-0.004$ \textit{eb}, and $-0.050$ \textit{eb} for $^{28}$Si, $^{30}$Si, and $^{32}$Si, respectively. The $2_2^+$ state in $^{28}$Si which is widely apart from $2_1^+$  turns out to be a part of prolate-structure, consistent with recent comprehensive shell model analysis \cite{dorian2024}. Whereas in the case of $^{30}$Si, the $2_2^+$ and $2_1^+$ states are significantly closer than in the remaining two isotopes as shown in Fig.~\ref{fig:SM_levels}. Also, $2_2^+$ state in $^{30}$Si lies within $\sim 0.4$ MeV of the $0_2^+$ state. While reaching to $^{32}$Si, the $2_2^+$ state appears again to be a prolate. In $^{30}$Si, the destructive interference of proton and neutron quadrupole moment contributions holds true for both the $2_1^+$ and $2_2^+$ states, due to the influence of $d_{3/2}-s_{1/2}$ mixing. In contrast, for the higher-angular momentum, the mixing of $d-$ orbitals again supports the constructive interference of involved proton and neutron $Q_s$ contributions. While in $^{28}$Si and $^{32}$Si, the interference of proton and neutron $Q_s$ contributions is again found to be constructive for both the first-and second-excited $2^+$ states.

\begin{table}
\caption{\label{tab:sm}
Average shell model occupancies of active proton and neutron orbitals for few low-lying states in even-even $^{28-32}$Si isotopes. $\langle L_p\rangle$, $\langle L_n\rangle$ exhibit orbital angular momentum contributions of protons and neutrons, respectively.}
\resizebox{\columnwidth}{!}{%
\begin{tabular}{|c|ccc|ccc|c|c|}
\hline
Nucleus & \multicolumn{3}{c|}{Protons} & \multicolumn{3}{c|}{Neutrons} & & \\
$J^\pi$ & $1d_{3/2}$ & $1d_{5/2}$ & $2s_{1/2}$ & $1d_{3/2}$ & $1d_{5/2}$ & $2s_{1/2}$ & $\langle L_p \rangle$ & $\langle L_n \rangle$ \\
\hline\\
$^{28}$Si &  \\
\hline
$0_1^+$ & 0.627 & 4.659 & 0.714 & 0.627 & 4.659 & 0.714 & & \\
$0_2^+$ & 0.565 & 4.707 & 0.728 & 0.565 & 4.707 & 0.728 & & \\
$2_1^+$ & 0.725 & 4.291 & 0.984 & 0.725 & 4.291 & 0.984 & 0.959 & 0.959 \\
$2_2^+$ & 0.684 & 4.352 & 0.964 & 0.684 & 4.352 & 0.964 & 0.984 & 0.984 \\
$4_1^+$ & 0.864 & 4.270 & 0.867 & 0.864 & 4.270 & 0.867 & 1.928 & 1.928 \\
$4_2^+$ & 0.801 & 4.443 & 0.756 & 0.801 & 4.443 & 0.756 & 1.935 & 1.935 \\
\hline\\
$^{30}$Si &  \\
\hline
$0_1^+$ & 0.374 & 5.093 & 0.533 & 1.147 & 5.486 & 1.367 & & \\
$0_2^+$ & 0.368 & 5.218 & 0.414 & 1.568 & 5.600 & 0.832 & & \\
$2_1^+$ & 0.383 & 4.937 & 0.680 & 1.382 & 5.423 & 1.195 & 0.626 & 1.309 \\
$2_2^+$ & 0.418 & 4.692 & 0.891 & 1.489 & 5.277 & 1.234 & 0.790 & 1.136 \\
$4_1^+$ & 0.555 & 4.701 & 0.744 & 1.479 & 5.114 & 1.407 & 1.544 & 2.315 \\
$4_2^+$ & 0.682 & 4.548 & 0.770 & 1.680 & 5.173 & 1.147 & 2.176 & 1.676 \\
\hline\\
$^{32}$Si & \\
\hline
$0_1^+$ & 0.221 & 5.349 & 0.431 & 2.546 & 5.842 & 1.612 & & \\
$0_2^+$ & 0.225 & 5.540 & 0.235 & 3.351 & 5.883 & 0.766 & & \\
$2_1^+$ & 0.215 & 5.186 & 0.599 & 2.616 & 5.839 & 1.544 & 0.601 & 1.412 \\
$2_2^+$ & 0.206 & 5.473 & 0.321 & 2.366 & 5.920 & 1.714 & 0.347 & 1.845 \\
$4_1^+$ & 0.274 & 4.772 & 0.954 & 2.779 & 5.563 & 1.659 & 1.495 & 2.387 \\
$4_2^+$ & 0.909 & 4.823 & 0.267 & 2.476 & 5.711 & 1.813 & 2.999 & 0.922 \\
\hline
\end{tabular}}
\end{table} 

A peculiar feature from average shell model occupancies is observed in the second-excited $0_2^+$ states; the average occupancy of neutron $2s_{1/2}$ orbital remains nearly unchanged across all three Si isotopes. Still, the $B(E2; 0_2^+ \rightarrow 2_1^+)$ changes rapidly when going from $^{28}$Si to $^{30}$Si, due to configurational changes for generating the $2_1^+$ state in both.
